\documentstyle[epsf]{lamuphys}
\makeatletter
\let\chapter\hid@chapter
\makeatother

\newcommand{\co}{\; \; ,}
\newcommand{\nn}{\nonumber \\}
\newcommand{\scs}{\co \;}
\newcommand{\per}{ \; .}

\newcommand{\sem}{\;\; ; \; }

\newcommand{\boldphi}{{\mbox {\boldmath $\pi$}}}

\newcommand{\ed}{\end{document}}
\newcommand{\be}{\begin{equation}}
\newcommand{\ee}{\end{equation}}
\newcommand{\bea}{\begin{eqnarray}}
\newcommand{\eea}{\end{eqnarray}}

\begin{document}
\pagenumbering{arabic}
\thispagestyle{empty}
\setcounter{page}{0}
\begin{flushright}
CERN--TH/97--320\\
BUTP--97/32\\
November 1997
\end{flushright}

\vfill

\begin{center}
{\LARGE \bf Aspects of Chiral Dynamics$^*$}\\[40pt]

\large
J\"urg Gasser\\[0.5cm]
CERN, Theory Division, 1211 Geneva, Switzerland and\\
Institut f\"ur Theoretische Physik, Universit\"at Bern\\
Sidlerstrasse 5, 3012 Bern, Switzerland \\[10pt]
\end{center}
\vfill

\begin{abstract}
I discuss several topics in chiral perturbation theory -- in particular,
 I recall  pecu\-larities
of the chiral  expansion in the baryon sector.
\end{abstract}

\vfill
\begin{center}
Plenary talk given at the Workshop on Chiral Dynamics 1997 \\[5pt]
Mainz, Germany, Sept. 1 -- 5, 1997 \\[5pt]
To appear in the Proceedings
\end{center}

\vfill
\noindent * Work supported in part by Schweizerischer Nationalfonds.
\newpage

\newpage
\title{Aspects of Chiral Dynamics}

\author{J.~Gasser}

\institute{
CERN, Theory Division, 1211 Geneva, Switzerland and\\
Institut f\"ur Theoretische Physik, Universit\"at Bern\\
Sidlerstrasse 5, 3012 Bern, Switzerland
}

\maketitle

\begin{abstract}
I discuss several topics in chiral perturbation theory -- in particular,
 I recall  pecu\-larities
of the chiral expansion in the baryon sector.
\end{abstract}

\section{Introduction}

In my talk  I first discussed the symmetry properties of the QCD
hamiltonian and  its ground  state. In particular,
I considered   flavour  (isospin and $SU(3)$) and chiral
symmetries in some detail. Here, I
followed closely the article  by  \cite{hlmass},
to which I refer the reader.
 Then,
I outlined the effective low--energy theory of QCD in the meson
and baryon sector and illustrated it with a few examples.
 There are  many review articles
 on chiral perturbation theory available on the
 market, see e.g. Bijnens et al. (1995), Ecker (1995a,b), Gasser
(1995), Leutwyler (1991,1994b) and  Mei\ss ner (1993).
 Here, I shall therefore  concentrate
   on some aspects of baryon chiral perturbation theory and illustrate why
 the low--energy expansion  is   rather involved in this case.

\section{Effective theory}
 The QCD
lagrangian can be replaced at low energies with
an effective lagrangian that is formulated  in terms of the
asymptotically
observable fields, see Weinberg (1979), Gasser and Leutwyler (1984,1985).
 This effective lagrangian reads for
processes with pions alone
 \bea
{\cal L}_M =\frac{F^2}{4}\langle \partial_\mu U \partial ^\mu U^\dagger +
M^2(U+U^\dagger)\rangle\per\nonumber
\eea
Here, the matrix field $U$ is an element of $SU(2)$,
and the symbol $\langle A \rangle$ denotes the trace of the
matrix $A$.
 In the
following, I use  the
parametrization
 \bea
U=\sigma +\frac{i\phi}{F}\sem\phi=\left(\begin{array}{cc}
\pi^0&\sqrt{2}\pi^+\\
\sqrt{2}\pi^-&-\pi^0\end{array}\right)\co
\sigma=\left[{\mbox{\bf 1}}-\phi^2/F^2\right]^{\frac{1}{2}}\co\nonumber
\eea
and the notation
\bea
\phi&=& \sum_{i=1}^3\tau^i\phi_i\scs
\boldphi=(\phi_1,\phi_2,\phi_3)\per\nonumber
\eea
The coupling constant $F\simeq 93$ MeV measures the strength of the $\pi\pi$
interaction, and the quantity $M^2$ denotes the square of the
  physical pion mass
(that I denote with $M_\pi$) at lowest order in an expansion in powers of
$1/F$, see below.
It is proportional to the light quark
masses $m_u,m_d$,
\bea
M^2=2\hat{m}B\scs \hat{m}=\frac{1}{2}(m_u+m_d)\co\nonumber
\eea
where $B$ itself is related to the quark condensate, see \cite{glan}.
Note that the quantity $M^2$ occurs not only in the kinetic term
of the pion lagrangian, but also in the interaction: it acts
both as a mass parameter and as a coupling constant.
 The
lagrangian
${\cal L}_M$ is called the "non--linear sigma--model lagrangian".
This name has led to some confusion in the
literature about the
meaning of the effective lagrangian:
one is not replacing QCD
with a "chiral model", as this procedure is often called.
To the contrary, ${\cal L}_M$ can be used to
calculate processes at low energies, with a result that is -- as shown by
\cite{hlann} -- identical  to the one
in QCD.

\begin{figure}[h]
\begin{center}
\mbox{\epsfxsize=3cm
       \epsfbox{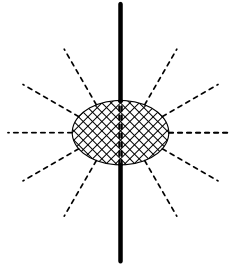}
     }
     \end{center}
 \caption{
The nucleon traveling through space, emitting and absorbing pions.
}
\label{fig1}
\end{figure}

In case we wish to consider also nucleons, one has to enlarge the above
lagrangian.
In the following, I will consider processes where a single baryon
(proton or neutron) travels in space, emitting and absorbing pions
in all possible ways allowed by chiral symmetry, see
 Fig. \ref{fig1}.
I do not consider processes with closed nucleon lines.
These contributions may be absorbed in a renormalization of the
coupling constants in the effective lagrangian
 \be
{\cal L}_{\,e\hspace{-.1mm}f\!f}={\cal L}_{M} +{\cal L}_{MB}\scs\label{eff1}
\ee
where the meson--nucleon interaction is given by
\bea
{\cal L}_{MB}&=&\bar{\Psi}
\left\{i\gamma^\mu \partial_\mu-m-\frac{g_A}{2F}
\gamma^\mu \gamma_5\partial_\mu\boldphi
 + O(\boldphi^2)\right\}
\Psi\per\label{eff2}
\eea
Here, $m$ is the nucleon mass in the chiral limit, and $g_A$ is the neutron
 decay constant $g_A\simeq 1.25$.
The
effective lagrangian (\ref{eff1}) contains the three
couplings $1/F,M^2$ and $g_A$.

\subsection{Tree level}
According to the rules set up in the sixties and seventies, one has simply to
evaluate tree graphs with
${\cal L}_{\,e\hspace{-.1mm}f\!f}$
to generate $S$--matrix elements
that are
in agreement with current algebra predictions. As is known today,
 this procedure
generates the leading order term in a systematic low--energy expansion of
the Green functions, see
\cite{weinberg} and \cite{hlann}.
 I illustrate it with two examples.

\subsubsection{The pion mass}
It suffices to consider the terms
in ${\cal L}_M$ that are quadratic in the pion fields,
\bea
{\cal L}_M =\frac{1}{2}\left\{\partial_\mu \boldphi \cdot \partial^\mu \boldphi
-M^2\boldphi^2\right\} +O(\boldphi^4)\per\nonumber
\eea
Therefore, the effective theory contains at tree level three
mass degene\-rate bo\-sons $\pi^+,\pi^-,\pi^0$, with
\bea
M_{\pi^\pm}^2=M_{\pi^0}^2= M^2\per\label{eqpionmass2}
\eea
At the leading order considered here, there is no isospin
splitting:
the masses of the charged and of the neutral pions are identical, see
 \cite{weinberg}.
A small mass
difference due to $m_u\neq m_d$ does show up only
 at next order in the chiral expansion.

\subsubsection{$\pi\pi$ scattering}
The full power of the effective lagrangian method comes into play when one
starts to evaluate scattering matrix elements. Consider for this purpose
elastic $\pi\pi$ scattering. The interaction part of the effective lagrangian
is
 \bea
{\cal
L}_{int}=\frac{1}{8F^2}\left\{\partial_\mu\boldphi^2\partial^\mu
\boldphi^2 -M^2(\boldphi\cdot\boldphi)^2\right\}
+O(\boldphi^6)\per\nonumber \eea
Since we calculate tree matrix elements, the terms at order O($\boldphi^6$) do
not contribute. The contributions with four
fields in the lagrangian contain two types of vertices: the first one has two
derivatives, while the second contains the parameter $M^2$ as a coupling
constant. In the following I consider the isospin symmetry limit $m_u=m_d$
and use the standard notation \bea
T^{ab;cd}=\delta^{ab;cd}A(s,t,u)
+\delta^{ac;bd}A(t,u,s)+\delta^{ad;bc}A(u,s,t)\nonumber
\eea
for the matrix element of the process
\bea
\pi^a(p_1)\pi^b(p_2)\rightarrow \pi^c(p_3)\pi^d(p_4)\co\nonumber
\eea
with the  Mandelstam variables
\bea
s=(p_1+p_2)^2\co t=(p_1-p_3)^2\co u=(p_1-p_4)^2\sem s+t+u=4M_\pi^2\per\nonumber
\eea
 The result of the calculation is
\bea
A
\stackrel{\mbox{\small{tree}}}{=}\frac{s-M^2}{F^2}
\stackrel{\mbox{\small{tree}}}{=}
\frac{s-M_\pi^2}{F_\pi^2}
\per\label{eqatree}
\eea
The second equal sign in Eq. (\ref{eqatree}) is based on the fact
 that the coupling $M^2$ can be replaced at tree level with
the square of the physical pion mass, see Eq. (\ref{eqpionmass2}),
and that the physical pion decay constant $F_\pi$ is equal to $F$
 in the same approximation. Of course, the result Eq. (\ref{eqatree}) agrees
 with the expression found by
\cite{weinpipi} using current algebra techniques.

In order to compare the above expression for the scattering matrix element with
the data, it is useful to consider the partial wave expansion of the amplitude.
For an illustration of this procedure, I refer the reader to the contribution
by \cite{ecker97}.

\subsection{Loops} As is well--known, unitarity requires that one considers
loops with the above effective lagrangian, see \cite{weinberg} -- tree
level results do not obey
the unitarity constraints for $S$--matrix elements.
I illustrate in the following chapter some features of loop contributions in
 the baryon sector.

 \section{Mass shifts -- relativistic framework}
\begin{figure}[h]
\begin{center}
\mbox{\epsfxsize=8cm
       \epsfbox{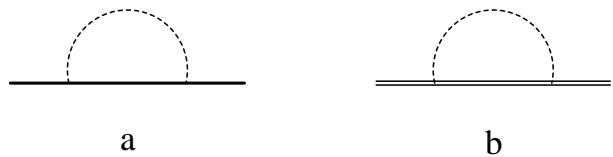}
     }
     \end{center}
 \caption{
Selfenergy graphs for a heavy particle. Fig. a: The solid (dashed) line denotes
the  propagator of the heavy (light) particle. Fig. b: The double line
indicates a
modified propagator for the heavy field. See text for details.
}
\label{fig2}
\end{figure}

To start with, we note that the interactions between the nucleon and the pions,
mediated through the effective lagrangian, will shift the value
of the nucleon mass $m$. In particular, as  the
coupling constant $M^2$ is proportional to the quark mass,
 the physical
nucleon mass will depend on $\hat{m}$ as well. So, I start
with a simple question: How does the nucleon mass depend on the
quark masses according to the effective lagrangian (\ref{eff1})?
At lowest order in the coupling $g_A$, we have to consider the
graph displayed in Fig. \ref{fig2}a,
 where the dashed line
denotes a pion with mass $M$, and the solid line
stands for the nucleon propagator. Note that this graph is of the type
considered in Fig. \ref{fig1}.

\subsubsection{The nucleon mass} The integral over  the meson momentum
in the graph Fig. \ref{fig2}a is ultraviolet
divergent. Regularizing this divergence by performing
the calculation in $d$ space--time dimensions,  the
shift becomes
\bea
\triangle
m=-mg(1+z)\frac{\Gamma(1-d/2)}
{(4\pi)^{d/2-2}}+O(1)\scs\nonumber
\eea
with
\bea
 g=\frac{3g_A^2m^2}{32\pi^2F^2}\scs
z=\frac{M^2}{m^2}\nonumber
 \eea
near the physical space--time dimension $d=4$.
 In order to
 eliminate this divergence, I introduce the counterterms
\bea
{\cal \delta L} =g(c_0
+c_1z)m\bar{\Psi}\Psi\nonumber\per
\eea
 Note that the
 structure of ${\cal \delta L}$ is different from the
original lagrangian (\ref{eff2}),
which thus corresponds to a non--renormalizable interaction.
 The result for the nucleon mass will  be
finite, provided that we tune the couplings $c_0,c_1$
appropriately as $d\rightarrow 4$. One obtains, see \cite{gss},
\bea
m_N&=&m\left[1+gh(z)\right]\scs\nn
 h(z)&=&\bar{c}_0+z(\bar{c}_1-1)-z\int_0^1\frac{ x
(2-x)\,dx}{x^2+z(1-x)}-z\ln{z}\per\label{mass1}
\eea
The quantities $\bar{c}_{0,1}$ denote renormalized,
scale independent coupling constants, independent of $M^2$.
 The exact relation to the $c_{1,2}$ introduced above
is of no relevance in the following, and I do therefore not display it here.

\subsubsection{Comparison with the pion mass}
\begin{figure}[h]
\begin{center}
\mbox{\epsfxsize=3cm
       \epsfbox{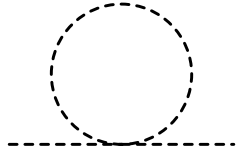}
     }
     \end{center}
 \caption{
Tadpole contribution to the pion  propagator. This graph generates the leading
correction to the pion mass in the chiral expansion.
}
\label{fig3}
\end{figure}

In order to discuss the special feature of the result Eq.
(\ref{mass1}), I also display the corresponding formula for the
shift of the pion mass, due to the graph Fig. \ref{fig3}.
Including the contribution from the counterterm $l_3$ in the effective
lagrangian at order $p^4$, one obtains at $m_u=m_d$
\bea
M_\pi^2=M^2\left\{1-\frac{M^2}{32\pi^2F^2}\bar{l}_3\right\}\scs\nonumber
\eea
where the renormalized coupling $\bar{l}_3$
depends
logarithmically on the quark mass,
\bea
M^2\frac{ d \bar{l}_3}{d M^2}=-1\per\nonumber
\eea

 The following comments are in order.
\begin{enumerate}
\item
The counterterms needed in the case of the nucleon mass are of
$O(1)$ and $O(p^2)$, whereas the tadpole contribution to the pion
mass requires a counterterm of $O(p^4)$. The fact that, in the
pion case, only this counterterm is required, is a feature of the
particular regularization scheme used. Had we introduced a
momentum cutoff, one would  have to add a counterterm of
order $p^2$  for the pion as well, see \cite{gaze}.
\item
The physical pion mass at one--loop order
 contains terms linear and
quadratic in the quark mass (up to the logarithm in $\bar{l}_3$).
On the other hand, the expression for the nucleon
mass as evaluated above  is a complicated
function of $\hat{m}$.
 Indeed,  expanding
the quantity $h(z)$ in Eq. (\ref{mass1}) around $z=0$ gives
\bea
m_N=m+mg\left\{\bar{c}_0+\bar{c}_1z-\pi z^{3/2}
-\frac{1}{2}z^2\ln{z} +\sum_{\nu=4}^{\infty} c_\nu
z^{\nu/2}\right\}\label{mass2}\per
\eea
\end{enumerate}

The origin of the different character of the one--loop expressions for
 the nucleon (pion)  mass
is easy to
identify: as the nucleon mass does not vanish in the chiral limit, it
provides an additional scale $m$ in the calculation -- besides $\hat{m}/F$,
one may also form $m/F$. It is obvious that this generates a problem
with the  chiral counting: in the meson case, loops
 contribute at a definite order in dimensional regularization. On the other
 hand, any power of the quark mass can be generated by chiral loops in the
 nucleon case. Below, I will illustrate how one can avoid this problem in
heavy baryon chiral perturbation theory (HBCHPT).

\subsubsection{Non--analytic terms} I now discuss the result Eq. (\ref{mass2})
in some detail.
First, consider the chiral limit.
It is convenient for the following  to normalize the
counterterm at order $p^0$ such that the nucleon mass stays at $m$
 when $\hat{m}\rightarrow 0$, i.e.,
$
\bar{c}_0=0$.
 Next, consider the term linear in the quark mass. It contains the
counterterm $\bar{c}_1$, which is related to the pion--nucleon
sigma--term, defined by
\bea
\sigma=\hat{m}\frac{\partial {m_N}}{\partial \hat{m}}\per\nonumber
\eea
{}From the above expression for the nucleon mass we obtain
\bea
\sigma
=mg\left\{\bar{c}_1z-\frac{3\pi}{2}z^{3/2}
-z^2\ln{z}+O(z^2)\right\}\per\nonumber
\eea
Therefore, the nucleon mass as well as the
sigma--term contain non--analytic contributions  of order $\hat{m}^{3/2}$
and $\hat{m}^2\ln{\hat{m}}$, see
\cite{gaze,gss}.
One may wonder what happens
to these terms once higher loop contributions are considered.
Of course, these  will start at order $p^0$
as well and give again rise to an infinite tower of terms.
 However, it can be shown  that the leading
non--analytic term in the expansion of the nucleon mass,
\bea
\delta m_N=-3\frac{g_AM^3}{32\pi F^2}\scs\label{nonanalytic}
\eea
is not touched by these contributions -- the coefficient of the term
proportional to $\hat{m}^{3/2}$ is fixed by chiral symmetry [\cite{gaze}],
 in contrast to the coefficient of the logarithmic singularity
$\sim \hat{m}^2\ln{\hat{m}}$, see
 \cite{glpr}.

 For an evaluation of all the
terms at
order $p^4$ in the
chiral expansion of the baryon octet,
 see \cite{bomei}, and the contribution of \cite{meissner97}
to these Proceedings.

\section{Non--relativistic formulation}

Heavy baryon chiral perturbation theory  is a quantum field theory
in which pure power counting for the baryons is restored, see \cite{jm}
and \cite{bkkm}: Each loop
generates exactly one term in the low--energy expansion of the quantity in
question. For example, in case of the nucleon mass, the graph
Fig. \ref{fig2}b generates the term (\ref{nonanalytic})
 and nothing else. In that graph, the double line denotes a properly
modified nucleon
propagator. I wish to illustrate in this section how this is achieved. In
 order to simplify the presentation, I consider the case of a scalar theory.
\subsubsection{Scalar theory}
Let
\bea
{\cal L}=\partial_\mu H^\dagger\partial^\mu H -m_H^2H^\dagger H+
\frac{1}{2}(\partial_\mu l \partial^\mu l -m_l^2l^2) +\kappa H^\dagger H l
\scs\nonumber
\eea
where $H$ ($l$) denotes a heavy (light) field of mass $m_H$ ($m_l$). The
shift in the heavy mass at lowest order in the expansion in the coupling
$\kappa$ is
due to the graph Fig. \ref{fig2}a, where the solid line now denotes the
propagator of the heavy scalar field $H$,
\bea
\delta m_H^2=i\kappa^2\int \frac{d^dl}{(2\pi)^d}\frac{1}{m_l^2-l^2}
\frac{1}{m_H^2-(p-l)^2}\scs p^2=m_H^2\per\label{heavymass}
\eea
Here, I have again regularized the expression by performing the integral
 in $d$
dimensions. Expanding the result in powers of the light mass gives
\bea
\delta m_H^2=\frac{\kappa^2}{16\pi^2}\left\{a_d +\pi\frac{m_l}{m_H}
+O(m_l^2\ln{m_l})\right\} \scs\nonumber
\eea
where $a_d$ is  independent of $m_l$ and contains a pole at
$d=4$, which is
removed by standard mass renormalization.
The next term illustrates that the shift in the mass contains a non--analytic
term of the square root type. This term can be picked out directly from
the original integral (\ref{heavymass}) in the following manner.

First, I consider the rest frame $p^\mu=(m_H,\vec{0})$, where
\bea
\delta m_H^2=i\frac{\kappa^2}{2m_H}\int\frac{d^dl}{(2\pi)^d}\frac{1}{m_l^2-l^2}
\frac{1}{l^0-l^2/2m_H}\per\nonumber
\eea
Now, in the large $m_H$ limit, I neglect the mass in the denominator of the
integrand and consider the integral
$$
J_m\doteq i\frac{\kappa^2}{2m_H}\int\frac{d^dl}{(2\pi)^d}\frac{1}{m_l-l^2}
\frac{1}{l^0}\scs
$$
which is linearly divergent. By performing the integral in $d$ dimensions,
 one finds that $J_m$ is finite at $d=4$,
\bea
J_m=\frac{\kappa^2}{16\pi}\frac{m_l}{m_H}\scs\label{massnon}
\eea
which is exactly the non--analytic term in the mass shift found above!
[One could as well introduce e.g. a momentum cutoff. The integral $J_m$ then
 contains, aside from the non--analytic piece (\ref{massnon}), a linear
divergent part which is independent of the light mass, and terms that vanish
as the cutoff is removed.]
Note that neglecting $l^2/2m_H$  in the denominator does not represent a
legal mathematical procedure: the result of the operation does not correspond
to the large $m_H$ expansion of the original integral -- on the other hand,
it does
correctly reproduce the leading mass correction, as we have just
seen.\footnote{Tang (1996) has proposed similar manipulations in ordinary
relativistic baryon chiral perturbation theory, recovering the results
of HBCHPT at one--loop order. See also Ellis and Tang (1997).}

HBCHPT is the science how to achieve these manipulations systematically
and legally in a lagrangian framework. Again, I illustrate it with the
 scalar theory.
\subsubsection{Non--relativistic formulation}
First, I replace the heavy field $H$ with a non--relativistic complex scalar
field $\Phi$,
\bea
{\cal L}\rightarrow {\cal L}_{NR}=\Phi^\dagger
\left(i\partial_t-\sqrt{m_H^2-\triangle}\right)\Phi +
\frac{1}{2}\left(\partial_\mu l \partial^\mu l-m_l^2 l^2\right)
+\frac{\kappa}{2m_H}\Phi^\dagger\Phi l \per\nonumber
\eea
The coupling constant has been adjusted in order to generate the correct
low--energy behavior of the tree amplitudes. Next, I expand the
non--local differential operator,
\bea
\sqrt{m_H^2-\triangle}=m_H-\frac{\triangle}{2m_H}+\cdots\scs\nonumber
\eea
and  put the derivative terms in the interaction,
\bea
{\cal L}_{NR}\!=\!\Phi^\dagger\!\left(i\partial_t-m_H\right)\!\Phi
\!+\!\frac{1}{2}\left(\partial_\mu l \partial^\mu l -m_l^2 l^2\right)
\!+\!\frac{\kappa}{2m_H}\Phi^\dagger\Phi l
+\Phi^\dagger\frac{\triangle}{2m_H}\Phi
 +\cdots\per\nonumber
\eea
The propagator of the non--relativistic field is $(p^0-m_H)^{-1}$ in Fourier
space. Dropping the terms with derivatives, the graph Fig.
 \ref{fig2}b gives
$$
\delta m_H^2=\frac{\kappa^2}{16\pi}\frac{m_l}{m_H}\scs
$$
and nothing else, which is exactly the needed result.

HBCHPT allows one to perform the low--energy expansion in the baryon sector
 (one external nucleon) in a systematic manner, by proceeding similarly to
  the scalar field just discussed. The chiral expansion of the
quantities evaluated earlier in the relativistic framework can then be
obtained  much easier  -- I refer the interested reader to the
review by \cite{chpt8}. In fact, an impressive
amount of calculations has been done in recent years e.g. by Bernard,
Kaiser and Mei\ss ner and others in this framework, see
 \cite{meissner97}, where also an outline of HBCHPT is presented.

\subsubsection{Comment}
There is one point that I wish to mention concerning this way of performing
 the chiral expansion. As I have just illustrated, HBCHPT is a clever
method to organize the calculations and to keep track of power counting. On
the other hand, the physics does, of course, not change. To illustrate,
consider e.g. the elastic pion--nucleon scattering amplitude. It has been
evaluated to one loop in the relativistic formulation some time ago by
 \cite{gss}.
 Expanding that result in powers of momenta and of quark masses, one would
obtain
\bea
A^{1 loop}=A_{1}+A_{2}+A_{3}+\cdots \hspace{1cm}
{\mbox{(relativistic framework)}}\scs\nonumber
\eea
where $A_n$ is of order $p^n$. I see no reason to doubt that the one--loop
calculation in HBCHPT, performed
recently by \cite{bkmpn} and by \cite{mojzis97a}, is identical to
\bea
A^{1 loop}=A_{1}+A_{2}+A_{3}\hspace{1cm}{\mbox{(HBCHPT)}}\per\nonumber
\eea
In this sense, the physics of HBCHPT is the same
 as the one of the original relativistic formulation.
On the other hand,
HBCHPT has the advantage that one is
certain to have collected all the terms at a given order in the chiral
expansion even for nonzero quark mass -- something that would be very
difficult to prove in the relativistic framework.

\section{Rate of convergence} Convergence of the chiral series is sometimes
very slow in the nucleon sector. To illustrate this, I consider the scalar
form factor of the nucleon,
\bea
\langle N(p')|\hat{m}(\bar{u}u+\bar{d}d)|N(p)\rangle=\bar{u}(p')u(p)\sigma(t)
\sem t=(p'-p)^2\per\nonumber
\eea
At zero  momentum transfer, the scalar form factor coincides with the
sigma--term considered above, $\sigma(0)=\sigma$. The difference
\bea
\triangle_\sigma=\sigma(2M_\pi^2)-\sigma(0)\nonumber
\eea
plays a central role in the extraction of the sigma--term from the
elastic pion--nucleon scattering amplitude.
The chiral expansion for this difference gives at leading order
\bea
\triangle_\sigma&=&\frac{3g_A^2M_\pi^3}{64\pi F_\pi^2}
\simeq 7.5\,\, {\mbox {MeV}}\hspace{1cm}({\mbox {leading order}})\scs\nonumber
\eea
 see \cite{glpr}.
On the other hand, a dispersive analysis -- that includes all orders in
 the quark mass expansion -- leads to
\bea
\triangle_\sigma=15\,\, {\mbox {MeV}}
\hspace{1cm}{\mbox{(dispersive analysis)}}\scs\label{mass3}
\eea
see \cite{gls91}. This example shows quite drastically that higher orders
in the quark mass
 expansion may be large -- even as large as the leading term, as the
present example shows. Indeed, by including the Delta resonance as an
explicit degree of freedom in the effective lagrangian, \cite{bkmz}
also find
$\triangle_\sigma \simeq 15$  MeV.
 The difference between this value and the leading order result
$\triangle_\sigma=7.5$ MeV is due to terms at order $p^4$ and higher. Of
course, if one would not know the result (\ref{mass3}) of the dispersive
 analysis, one
could only conclude form their calculation that there are potentially
large corrections
to the leading order result -- and nothing more. Whether the remaining
terms at order $p^4$ or even higher order
contributions are large cannot be decided from this one--loop calculation.
 To pin  them down in a purely chiral expansion framework
is very difficult -- in this case, the dispersive analysis is
 more efficient.

As this example  illustrates, the Delta degree of freedom may generate
large perturbations. Hemmert, Holstein and Kambor (1997)
 have therefore developed a
framework where the Delta resonance is taken into account in a systematic
manner.
One counts
the pion mass, as well as the difference between the
 Delta and the nucleon mass,  as
quantities of order $\epsilon$. For example,
the ratio
\bea
 \frac{M_\pi^3}{M_\pi+m_{\small{Delta}}-m_N}\nonumber
\eea
is then considered as order $\epsilon^2$, whereas it is
  order $p^3$ in conventional
power counting.
 For details concerning this framework,
I refer the reader to the contribution of \cite{kambor97}.

There are several reasons for the slow convergence of the chiral expansion in
the nucleon sector. First, as we have just seen, the proximity of the
Delta resonance may cause large corrections. Although there is a mass gap
between the Delta and the nucleon also in the chiral limit, the Delta does
stay nearby in the real world and cause large effects through small energy
denominators. Second, the ratio $M_\pi/M_N\simeq 1/7$, in which the
amplitudes are expanded, is not that small. Third and most importantly in my
opinion, the chiral expansion of e.g. the full elastic pion--nucleon scattering
 amplitude
is of the form
$$
A=A_{1}+A_{2}+A_{3}+A_{4}+\cdots\scs\nn
$$
i.e., there is a chain of even and odd powers
 in the momenta. In each chain, one needs at least the leading and the
next--to--leading order term to have a reliable prediction.
 I see no
reason to trust any calculation that does not include all these terms
 -- only in this case can one check to some extent whether one
has obtained a satisfactory approximation. This means that we need the
terms of order $p^4$ in the pion--nucleon amplitude. For a discussion of
the results at order $p^3$, see
Moj\v zi\v s (1997a,b)
 and
 \cite{ecker97}.

In fact, the calculation of the terms at order $p^4$ in the baryon mass,
 recently carried out by \cite{bomei}, allows for such a check.
I consider the mass of the $\Xi$ and write again
\bea
m_\Xi=m_0+m_2+m_3+m_4+\cdots\per\nonumber
\eea
According to these authors, the first chain reads
$$
(m_0,m_3)=(770,-893) \hspace{2mm}{\mbox{MeV}}\scs
$$
whereas
$$(m_2,m_4)=(847, 600) \hspace{2mm}{\mbox{MeV}}\per
$$
Since $m_3 (m_4)$ should be a correction to $m_0 (m_2)$, I consider this a
disaster for the chiral expansion. For a different opinion,
see \cite{bomei}, and the contribution of \cite{meissner97} to this workshop.

Note that, in the meson sector, the situation is very much different:
 The effective action contains only even powers of the momenta -- a one--loop
 calculation therefore often suffices in the case where the leading order
 term starts at tree level.

\section{Mass effects in the low--energy constants}

There is one more feature of chiral expansions that one can nicely illustrate
with
the nucleon mass and the sigma--term, that  I write as
\bea
m_N&=&m+\sigma+\frac{\pi}{2}mgz^{3/2}+O(z^2\ln{z})\co\nn
\sigma&=&m g\bar{c}_1z -\frac{3\pi}{2}mg z^{3/2}
 +O(z^2\ln{z})\per\nonumber
\eea
These expressions contain the two low--energy constants $m$
 and ${\bar{c}}_1$, which are not determined by chiral symmetry.
 As is usual, one may rely on  experimental information to pin them down.
 I illustrate the procedure in the following, using
$$
(m_N,\sigma,F,M)=(940,45,93,135)\,\,\, {\mbox{MeV}}\sem g_A=1.25.
$$
At leading order in the chiral expansion, one has
$$
m_N=m\scs\sigma=0\,\,\Rightarrow m=940 \,\,{\mbox{MeV}}\per
$$
At next order,
$$
m_N=m+\sigma\scs\sigma=mg\bar{c}_1z\,\,\Rightarrow m=895 \,\,{\mbox{MeV}}\scs
\bar{c}_1=1.6\per
$$
Finally, from the expressions at order $p^3$, I find
$$
m=888 \,\,{\mbox{MeV}}\scs\bar{c}_1=2.3\per
$$
The fact that the values of the low--energy constants depend on the
 order we are considering seems to be in contradiction with calling them
``constants''. Of course, these quantities indeed are
 quark mass independent. However, once we determine them from data,
one is using a specific order in the chiral expansion,
whereas the data do include  the quark mass effects to all orders.
Some of these are therefore effectively absorbed in the low--energy
constants, as
 a result of which
one is  faced with a systematic uncertainty in the determination of
their values,
 even with infinitely precise data, as the above chain
\bea
m&=&940 \,\,{\mbox{MeV}}\rightarrow
  895 \,\,{\mbox{MeV}}\rightarrow
  888 \,\,{\mbox{MeV}}\scs\nn
\bar{c}_1&=&1.6\rightarrow 2.3\scs\nonumber
\eea
 nicely illustrates.

There is, on the other hand, at least in
principle a possibility to generate
 data without quark mass effects: lattice
calculations. Indeed, once it will be possible to e.g. determine the
value of the nucleon mass in the chiral limit from lattice simulations, we
 may  simply take that value for $m$. The other parameter at hand,
${\bar{c}}_1$, can be obtained by evaluating the derivative of the nucleon
mass with respect to the quark mass in the chiral limit.
Needless to say that these are very difficult quantities to measure on the
lattice.

\section*{Acknowledgements}
I thank Alex Gall for enjoyable discussions concerning the
material in section four,
 and  Gerhard Ecker, Joachim Kambor,
 Ulf Mei\ss ner and
Martin
Moj\v zi\v s for
 discussions on
   atrocities in baryon chiral perturbation theory.
  Furthermore, I thank Aron Bernstein, Dieter
Drechsel and Thomas Walcher
 for the efficient organization of this Workshop.

 \end{document}